\documentclass{PoS}

\title{Revealing the nature of unidentified {\it INTEGRAL} sources through optical spectroscopy: an overview}

\ShortTitle{An overview on the nature of unidentified {\it INTEGRAL} sources}

\author{\speaker{Pietro Parisi}\\
        INAF -- Istituto di Astrofisica Spaziale e Fisica Cosmica di Bologna, Via Gobetti 101, I-40129 Bologna, Italy\\
        and Dipartimento di Astronomia, Universit\`a di Bologna, Via Ranzani 1,
I-40129 Bologna, Italy\\
        E-mail: \email{parisi@iasfbo.inaf.it}}

\author{{Nicola Masetti}\\
        INAF -- Istituto di Astrofisica Spaziale e Fisica Cosmica di Bologna, Via Gobetti 101, I-40129 Bologna, Italy.\\
        E-mail: \email{masetti@iasfbo.inaf.it}}

\author{on behalf of the IBIS survey team}

\abstract{Thanks to the {\it INTEGRAL} satellite the way of looking at the hard X-ray sky above 20 keV has \
changed substantially. Through the unique imaging and spectroscopy capabilities of the IBIS instrument that has formed the basis \ 
of the {\it INTEGRAL} surveys, this satellite has improved the knowledge on hard X-ray sources in terms of sensitivity and \
positional accuracy. Many of the sources belonging to these surveys are however of unidentified nature, but the \ 
combined use of available information at longer wavelengths (mainly soft X-rays and radio) and above all optical spectroscopy\ 
on the putative counterparts of these hard X-ray objects can reveal their exact nature. Since 2004 our group identified \ 
more than 100 {\it INTEGRAL} sources, reducing drastically the percentage of unidentified objects in the various IBIS surveys \
and allowing statistical studies on them. Here we present a summary of this identification work and an outlook of our \
preliminary results on identification of newly-discovered sources belonging to the 4$^{\rm th}$ IBIS catalog.}

\FullConference{The Extreme sky: Sampling the Universe above 10 keV - extremesky2009,\\
		October 13-17, 2009\\
		Otranto (Lecce) Italy}

\begin{document}
\section{Introduction}
\vspace{-0.3cm}
Since is launch in 2002 the {\it INTEGRAL} (Winkler et al. 2003) satellite is performing four main scientific aims: (i) a deep 
exposure of the Galactic central radian, (ii) regular scans of the Galactic Plane, (iii) pointed observations of the Vela region 
and (iv) Target of Opportunity follow-ups. IBIS (Ubertini et al. 2003) is a hard X-ray imaging instrument onboard {\it INTEGRAL} 
with a large field of view (30$^{\circ}$), and it is the basis of several {\it INTEGRAL} surveys. Through this unique 
capabilities, IBIS 
permits the detection of sources at the mCrab level with a typical localization accuracy of 2-3 arcmin above 20 keV (Gros et al. 
2003). From the first to the fourth IBIS survey catalog, both sensitivity and sky coverage improved substantially, enabling the 
increase of the number of detected hard X-ray sources from 123 in the 1$^{\rm st}$ catalog to 723 in the 4$^{\rm th}$ one. A 
fraction of these objects ($\sim$30\% in all catalogs) had no known or evident counterpart at other wavelengths and therefore 
could not be associated with any known class of high-energy emitting sources. For this reason, since 2004 our group has been 
performing an observational campaign employing telescopes located in the northern and the southern hemispheres to obtain optical 
spectroscopy of the putative counterparts of these hard X-ray emitting objects in order to determine their actual nature.

Here we want to briefly illustrate the method we use to associate the optical counterpart to the corresponding unidentified 
hard X-ray source and the progress of our identifications from the 1$^{\rm st}$ to the 4$^{\rm th}$ IBIS survey catalog by
pointing out the contribution of our group to this identification work.
\vspace{-0.3cm}
\section{The IBIS/{\it INTEGRAL} soft gamma-ray surveys}
\vspace{-0.3cm}
The 1$^{\rm st}$ IBIS soft gamma-ray survey catalog was performed in the first year of satellite operations, the regular scans 
of the Galactic Plane yielding a survey with a sensitivity down to $\sim$1 mCrab (Bird et al. 2004). This allowed detecting 
more than 120 sources (22\% of which were unidentified), many of them detected for the first time above 20 keV. 
The second IBIS survey catalog (Bird et al. 2006) increased the sensitivity unveiling more than 200 sources with $\sim$27\% 
of them unidentified.
Within the third IBIS survey catalog (Bird et al. 2007) 421 hard X-ray sources were detected, and $\sim$28\% had no 
classification, while the fourth catalog (Bird et al. 2010) contains 723 hard X-ray emitting objects, with as much as $\sim$ 
29\% of unidentified sources.

\begin{figure*}[th!]
\begin{center}
\includegraphics[width=12cm]{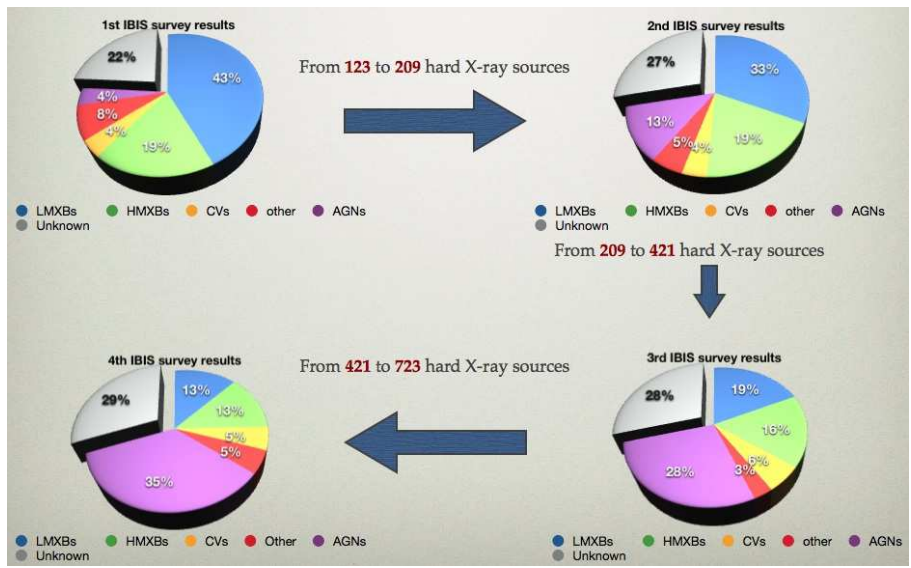}
\caption{Evolution of the percentages of the detected X-ray sources (divided according to their classification) from the 
1$^{st}$ to the 4$^{th}$ IBIS survey catalog.
}\label{sur}
\end{center}
\end{figure*}

As one can see in Fig. \ref{sur}, the majority of the identified sources in the first three catalogs was made of Galactic 
objects, while in the 4$^{\rm th}$ one extragalactic sources constitute the largest group. Moreover, a marked evolution in 
the percentages of Galactic sources is apparent: Low Mass X-ray Binaries (LMXBs) fell from 43\% of the
1$^{\rm st}$ survey to 13\% in the 4$^{\rm th}$ one; a similar, albeit less sharp, reduction holds for High Mass X-ray Binaries 
(HMXBs). The number of detected Cataclysmic Variables (CVs) kept instead stable across the four catalogs. 

As already mentioned, the detection of extragalactic sources, that is Active Galactic Nuclei (AGNs), skyrocketed from 4\% to 
35\%. Besides, 
while the number of detected sources increased dramatically (by a factor of $\sim$6) across the four surveys (of course due to 
the larger and larger instrument exposure available), the percentage of the unknown objects remained almost constant. As we 
will show below, these two results are a clear demonstration that our observational multisite campaign for the search of 
optical counterparts of {\it INTEGRAL} hard X-ray sources is highly effective. 
\vspace{-0.3cm}
\section{The identification method}
\vspace{-0.3cm}\
The first step for the determination of the nature of unidentified sources belonging to the IBIS catalogs is to search for 
counterparts at other wavelengths. This, albeit the {\it INTEGRAL} error boxes are definitely smaller and more easily explorable 
with respect to those afforded by past hard X-ray missions, is still quite complicated due to the fact that the uncertainty 
circle area (which is of the 
order of $\sim$10 arcmin$^2$) contains a large number of objects, especially at optical wavelengths. One therefore needs to 
reduce the search area down to a size of less than a few arcseconds.

Cross correlations with catalogues in other wavebands can then be used to reduce the positional uncertainty to facilitate the 
identification process. Stephen et al. (2006), by cross-correlating the 2$^{\rm nd}$ IBIS and the {\it ROSAT} catalogs, 
demonstrated that when a bright, single soft X-ray object is found within the IBIS error circle, it is almost certainly the 
lower-energy counterpart of the {\it INTEGRAL} source. Thus, the presence of a catalogued or archival {\it Swift}, {\it ROSAT}, 
{\it Chandra} and/or {\it XMM-Newton} source within the IBIS error circle of a hard X-ray object marks the position of its 
longer wavelength counterpart with a precision of a few arcseconds or better. Thereby reducing the search area of a factor of 10$^{4}$.

Similarly, when no soft X-ray information is available, far-infrared or radio catalogs can be used for this task.
In this case, however, the reliability of the correlation is less strong (see e.g. Masetti et al. 2008a and references 
therein), and one always needs confirmation by means of a clear soft X-ray detection from the putative counterpart.
Likewise, the presence of a bright and peculiar optical object (e.g. an unidentified galaxy, or a poorly studied 
emission-line star) within the IBIS error circle can be a hint for it to be the counterpart of the IBIS source; again, however, 
one has to wait for confirmation of emission at soft X-rays from the optical object to definitely prove any association.

Once this smaller, arcsec-sized position is available, a search for optical counterparts within its area is made with the use 
of archival images (e.g., those of the DSS II-red survey\footnote{{\tt http://archive.eso.org/dss/dss}}). Then, optical 
spectroscopy is performed on the optical object(s) within this smaller error box: the source which presents remarkable spectral 
features (typically strong Balmer, Helium or forbidden emission lines) can definitely be identified as the optical counterpart
of the IBIS hard X-ray emitter. The characteristics of the optical spectrum eventually allow us to determine distance, 
reddening, chemical composition, and in the very end the nature, of the considered {\it INTEGRAL} unidentified source.
\vspace{-0.3cm}
\section{Telescopes involved in this activity}
\vspace{-0.3cm}
Starting with the very first pilot project performed at the Loiano Telescope (Masetti et al. 2004),
during the more than 5-year long hunt for the identification of {\it INTEGRAL} sources set on by our group, optical 
spectroscopy useful for our identification was performed at various telescopes worldwide. Below we present the list of all 
ground-based facilities used until now within this project:

\begin{itemize}
\vspace{-0.1cm}
\item 1.5m telescope at the Cerro Tololo Interamerican Observatory, Chile;
\vspace{-0.3cm}
\item 1.52m ``Cassini'' telescope of the Astronomical Observatory of Bologna, in Loiano, Italy;
\vspace{-0.3cm} 
\item 1.8m ``Copernicus'' telescope at the Astrophysical Observatory of Asiago, in Asiago, Italy;
\vspace{-0.3cm}
\item 1.9m ``Radcliffe'' telescope at the South African Astronomical Observatory, in Sutherland, South Africa;
\vspace{-0.3cm}
\item 2.1m telescope of the Observatorio Astr\'onomico Nacional in San Pedro M\'artir, Mexico;
\vspace{-0.3cm}
\item 2.15m ``Jorge Sahade'' telescope at the Complejo Astron\'omico el Leoncito, Argentina;
\vspace{-0.3cm}
\item 3.58m telescope ``Telescopio Nazionale Galileo'' at the Observatorio of the la Roque de Los Muchachos 
in Canary Islands, Spain; 
\vspace{-0.3cm}
\item 3.58m NTT of the ESO Observatory in La Silla, Chile;
\vspace{-0.3cm}
\item 3.6m telescope of the ESO Observatory in La Silla, Chile;
\vspace{-0.3cm}
\item 4.2m ``William Herschel Telescope'' at the Observatory of Roque the Los Muchachos in Canary Islands, Spain;
\vspace{-0.3cm}
\end{itemize}
plus archival spectra from 6dF\footnote{{\tt http://www.aao.gov.au/local/www/6dF}} and SDSS\footnote{{\tt http://www.sdss.org}}.

\section{Results}

Up to the time of this conference (October 2009), within our identification program we produced 9 refereed papers 
(Masetti et al. 2004, 2006a,b,c,d, 2007, 2008a,b, 2009) plus a number of conference proceedings and of short 
communications. Going into details, our work allowed us to spectroscopically determine or confirm the nature of 104 
unidentified {\it INTEGRAL} sources, which can be divided into several subclasses as follows (see also Fig. \ref{res}):

\begin{itemize}
\vspace{-0.3cm}
\item 4 (persistent) LMXBs;
\vspace{-0.3cm}
\item 14 Be/X HMXBs (often with a highly reddened optical counterpart);
\vspace{-0.3cm}
\item 5 HMXBs with supergiant companion (often fast X-ray transients);
\vspace{-0.3cm}
\item 57 nearby AGNs (30 Seyfert 1 and 27 Seyfert 2) with redshift between 0.011 and 0.422;
\vspace{-0.3cm}
\item 3 X-ray Bright, Optically Normal Galaxies (XBONGs);
\vspace{-0.3cm}
\item 2 high-$z$ blazars (with redshift $\geq$1);
\vspace{-0.3cm}
\item 2 BL Lacs;
\vspace{-0.3cm}
\item 12 magnetic CVs;
\vspace{-0.3cm}
\item 4 Symbiotic stars;
\vspace{-0.3cm}
\item 1 Active star.
\vspace{-0.3cm}
\end{itemize}

Among these, we would like to mention a few outstanding objects, such as the Symbiotic X-ray Binary IGR J16194$-$2810
(Masetti et al. 2007) and the high-redshift ($z$ = 2.40) blazar Swift J1656.3$-$3302 (Masetti et al. 2008a).

We also stress that this work allowed the detection of a large number of new AGNs, especially in the so-called `Zone of 
Avoidance', i.e. along the Galactic Plane, where the presence of Galactic dust and neutral hydrogen severely hampered
past studies of AGNs at both optical and soft X-ray wavelengths. It also gave us the possibility of detecting a 
substantial number of new, possibly magnetic, CVs (e.g. Landi et al. 2009; Scaringi et al. 2009). 

It is moreover remarked that, despite recent claims (Cerutti et al. 2009), the present program halved the number of 
unidentified sources detected in the 3$^{\rm rd}$ IBIS catalog, and the same bright goal is expected for the 4$^{\rm th}$ 
survey (see next Section).

Moreover, as a service to the community, we regularly maintain a web archive reporting the main properties of each 
{\it INTEGRAL} source identified through optical and near-infrared spectroscopy. This archive can be found at the URL:

\begin{figure*}[th!]
\begin{center}
\includegraphics[width=5cm]{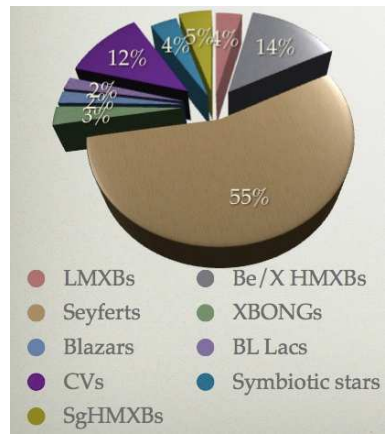}
\caption{The percentages of the various unidentified sources identified by our group up to October 2009, divided 
according to their classification.}
\label{res}
\end{center}
\end{figure*}

\noindent
{\tt http://www.iasfbo.inaf.it/extras/IGR/main.html} .
\vspace{-0.3cm}
\section{Outlook}
\vspace{-0.3cm}
As already remarked, the 4$^{\rm th}$ IBIS survey has about 29\% of sources which lack an obvious counterpart,
which means that it hosts 208 sources of unidentified nature.
As for the past catalogs, we already started the identification work for these sources by means of optical spectroscopy, 
and we alredy selected a sample of 25 sources for which a classification could be achieved using the approach 
illustrated above {(Masetti et al. 2010)}. The majority of these newly-identified sources are AGNs (68\%), followed 
by CVs and X-ray binaries (both 16\%). This new lot of identified sources already reduced by $\sim$ 12\% the whole amount 
of unidentified sources in the 4$^{\rm th}$ IBIS catalog.

In conclusion, we point out that we are running a similar project in relation with the {\it Swift/BAT} sources belonging to 
different catalogs (see Landi et al. 2007 and Parisi et al. 2009): we could identify or better classify 28 hard 
X-ray emitting AGNs. Likewise, we are also performing a similar work on the unidentified sources detected with the 
{\it Fermi} satellite: this allowed us to classify the GeV source 0FGL J2001.0+4352 as a BL Lac object (Bassani et al. 
2009).

\end{document}